# Network Dimensions in the Getty Provenance Index

Maximilian Schich[i, ii*], Christian Huemer[iii], Piotr Adamczyk[iv], Lev Manovich[v], and Yang-Yu Liu[vi,vii]

In this article we make a case for a systematic application of complex network science to study art market history and more general collection dynamics. We reveal social, temporal, spatial, and conceptual network dimensions, i.e. network node and link types, previously implicit in the Getty Provenance Index® (GPI).[1] As a pioneering art history database active since the 1980s, the GPI provides online access to source material relevant for research in the history of collecting and art markets. Based on a subset of the GPI, we characterize an aggregate of more than 267,000 sales transactions connected to roughly 22,000 actors in four countries over 20 years at daily resolution from 1801 to 1820. Striving towards a deeper understanding on multiple levels we disambiguate social dynamics of buying, brokering, and selling, while observing a general broadening of the market, where large collections are split into smaller lots. Temporally, we find annual market cycles that are shifted by country and obviously favor international exchange. Spatially, we differentiate near-monopolies from regions driven by competing sub-centers, while uncovering asymmetries of international market flux. Conceptually, we track dynamics of artist attribution that clearly behave like product categories in a very slow supermarket. Taken together, we introduce a number of meaningful network perspectives dealing with historical art auction data, beyond the analysis of social networks within a single market region. The results presented here have inspired a Linked Open Data conversion of the GPI, which is currently in process and will allow further analysis by a broad set of researchers.[2]

**Background**

In recent years, quantitative and computational research in art history have made and continue to make essential advances, harnessing an increasing availability of open data, adopting and contributing to ever more sophisticated methods of data science and visualization. Together with traditional qualitative inquiry, such research addresses "ideographic" specifics, where each historical source or event is considered unique, as well as more general "organized complexity" or "nomothetic" patterns, where larger amounts of evidence combine to non-random macroscopic structures or even physical laws.[3] Provenance research in general and research on the Getty Provenance Index (GPI) in particular are no exception to this development (cf. a number of contributions in press to appear in a volume at Getty publications). Both timely and late, this article provides citable proof that the GPI can be seen as a time-dependent network of complex networks, with multiple node and link types, subject to non-trivial dynamics that warrant both a deeper qualitative look and quantitative modeling in the sense of network and complexity science.[4]

The article is timely, as related topics currently emerge to become mainstream in both art research and the multidisciplinary science of complex networks. When we started our project in 2010, there was very little work on what was then called "interdependent networks" in general network science[5], and "networks of complex networks" in art research[6] in particular. General network science, back then, started to combine analyses of interdependent networks such as the internet and the power grid. Beside this strain of research, the study of networks of complex networks is also rooted in so-called "knowledge graphs"[7], i.e. databases typically including large numbers of node and link types, where each link type constitutes a complex network with characteristic individual properties, connected to a meta-network of networks by a data model definition. Since 2010, the general network science of now so-called "time-dependent" or "temporal, and "multiplex" or "multilayer" networks have grown to a full and rapidly growing body of literature that is subject to its own review papers.[8] Furthermore, when we started our work, the GPI data was proprietary and subject to a

---

**Affiliations**

[i]School of Arts, Technology, and Emerging Communication, The University of Texas at Dallas, Richardson/TX. [ii]The Edith O'Donnell Institute for Art History, The University of Texas at Dallas, Richardson/TX. [iii]Getty Research Institute, Los Angeles/CA. [iv]ITHAKA, New York/NY. [v]The Graduate Center, City University of New York, New York/NY. [vi]Channing Division of Network Medicine, Brigham and Women's Hospital and Harvard Medical School, Boston/MA. [vii]Center for Cancer Systems Biology, Dana Farber Cancer Institute, Boston/MA. * first and corresponding author (maximilian@schich.info)

**Footnotes**

[1] cf. http://www.getty.edu/research/tools/provenance/, accessed June 7, 2017
[2] Selected GPI datasets, to be superseded by the Linked Open Data release are available at https://github.com/gettyopendata/provenance-index-csv
[3] cf. Schich, Maximilian. „Figuring out Art History." *International Journal for Digital Art History* 2 (2016); and Manovich, Lev. „Data science and digital art history." *International Journal for Digital Art History* 1 (2015).
[4] For an introduction into network and complexity science in general see Barabási, Albert-László. *Network science*. Cambridge University Press, 2016; and Mitchell, Melanie. *Complexity: A guided tour*. Oxford University Press, 2009.
[5] Leicht, E. A., and Raissa M. D'Souza. „Percolation on interacting networks." *arXiv preprint* arXiv:0907.0894 (2009); Buldyrev, Sergey V., Roni Parshani, Gerald Paul, H. Eugene Stanley, and Shlomo Havlin. „Catastrophic cascade of failures in interdependent networks." *Nature* 464, no. 7291 (2010): 1025-1028.
[6] Schich, Maximilian. „Revealing Matrices." in *Beautiful Visualization*, pp. 227-254. O'Reilly, 2010; Schich, Maximilian. „Netzwerke von komplexen Netzwerken in der (Kunst) Wissenschaft." In *Die Dynamik sozialer und sprachlicher Netzwerke*, pp. 161-178. Springer Fachmedien Wiesbaden, 2013.
[7] cf. Bizer, Christian, Tom Heath, and Tim Berners-Lee. „Linked data-the story so far." *Semantic services, interoperability and web applications: emerging concepts* (2009): 205-227; and Vrandečić, Denny, and Markus Krötzsch. „Wikidata: a free collaborative knowledgebase." *Communications of the ACM* 57, no. 10 (2014): 78-85.
[8] Cf. Boccaletti, Stefano, Ginestra Bianconi, Regino Criado, Charo I. Del Genio, Jesús Gómez-Gardenes, Miguel Romance, Irene Sendina-Nadal, Zhen Wang, and Massimiliano Zanin. „The structure and dynamics of multilayer networks." *Physics Reports* 544, no. 1 (2014): 1-122; Kivelä, Mikko, Alex Arenas, Marc Barthelemy, James P. Gleeson, Yamir Moreno, and Mason A. Porter. „Multilayer networks." *Journal of complex networks* 2, no. 3 (2014): 203-271; Holme, Petter, and Jari Saramäki. „Temporal networks." *Physics reports* 519, no. 3 (2012): 97-125.

non-disclosure agreement for the purpose of analysis beyond the standard query interface. The GPI also relied on a mostly flat file structure, returning lists of sales transactions for qualitative research purposes. Since we started our work, quantitative and computational research in art history have seen the emergence of relevant funding and publication venues in "digital humanities" and "cultural analytics". The topic of "Arts, Humanities, and Complex Networks" has been a subject of numerous contributions from a great variety of disciplines.[9] Finally, our contribution feeds into literature dealing with the complexity of auctions and markets more general.[10]

The project kicked off in June 2010 in a structured five-day discussion at the Getty Research Institute, where Maximilian Schich, Christian Huemer, Lev Manovich, and Piotr Adamczyk "imagineered" a set of possible figures and plots, that would reveal hidden aspects of the GPI. The project then continued on the sidelines, as Piotr Adamczyk was contributing to the "Google Art Project"[11], Lev Manovich was busy to establish "Cultural Analytics"[12], and Maximilian Schich was working towards "A Network Framework of Cultural History" to be published in Science Magazine.[13] Eventually Maximilian Schich invested several months of data preparation and analysis, with results, although never formally published, being used as a consulting instrument to make a strong case for a more systematic network science approach dealing with provenance research and art history in general. Presented at several occasions, including at the National Gallery, London in June 2013, at the International Conference on Network Science in Berkeley (NetSci2014), and CASVA Washington in November 2014, the results fostered a complete conceptual and technical overhaul of the GPI. With the 2016-2019 Provenance Index Remodel Project, the GPI databases will be transformed into Linked Open Data, an event-based model that will allow researchers to take more fully advantage of the unique scale and complexity afforded by data-driven methodologies.

**Network Dimensions in the GPI**
We originally received the GPI data, as analyzed in this article, from the Getty Research Institute in the form of four tab-separated text files, covering "British", "French", "Belgian", and "Dutch" sales respectively. The period from 1801 to 1820 did lend itself to data-driven analysis of the auction market as this subset of 267,661 records from four European countries is quite complete and clean. According to Getty experts, it is complete, in the sense that every painting transaction recorded in an auction catalog surviving to this day should be in the dataset. It is quite clean as these GPI records have been produced for print publication many years ago, which required to follow a number of editorial standards. Despite the availability of GPI data in print and digital format, a data-driven approach to the study of early 19th-century art markets in Europe remains a desideratum.[14]

To enable our analysis of the GPI subset, we have minimally processed the data to harmonize slightly inconsistent conventions of data entry, such as the use of commas or semicolons by various editors working on British or French sales. In a more major step within data preparation, we have then assigned unique individual IDs to so-far unidentified nodes within so-far implicit network dimensions. This includes sales-events, actors, dates, locations, artist attributions, artist nationalities, and bibliographic references. Beyond assigning intuitively different node types to axiomatically distinct dimensions, we have also separated social actors from artists as they serve a different, non-social, more conceptual function within the GPI, as explained further below.

Figure 1A shows the resulting GPI data model as a weighted node-link diagram.[15] The figure is equivalent to a classic entity-relationship model in database planning[16], only here with the node size indicating the number of existing nodes per node type, and the line width indicating the number of links per link type. The dotted lines indicate further induced link types, which we have constructed by shortcutting the link between actors and locations, effectively treating the sales event as a link. The GPI data model diagram in Fig. 1A makes clear that the GPI is first and foremost not about known objects, i.e. works of art. Instead, the GPI primarily deals with socio-economic interactions between actors in the roles of sellers, experts, auction houses, commissaires priseurs[17] (i.e. French government officials conducting auctions), and buyers. The thinner line for buyers as opposed to sellers indicates that not all recorded transactions are complete sales, where the offered object changes the owner. Moreover, buyer names in the data derive from handwritten annotations on the margins of some catalog copies and are usually not comprehensive. Beyond the social dimension, the diagram indicates that every recorded sales transaction has a sales location, a sales date, and at least one artist attribution with a further specified nationality, both of which can of course be "unknown" or "unspecified". Finally, the diagram shows a number of references to previous and subsequent events, via the previous sale, post sale, previous owner, and post owner link types, all of which are rather sparse for quantification, but offer interesting clues for qualitative research.

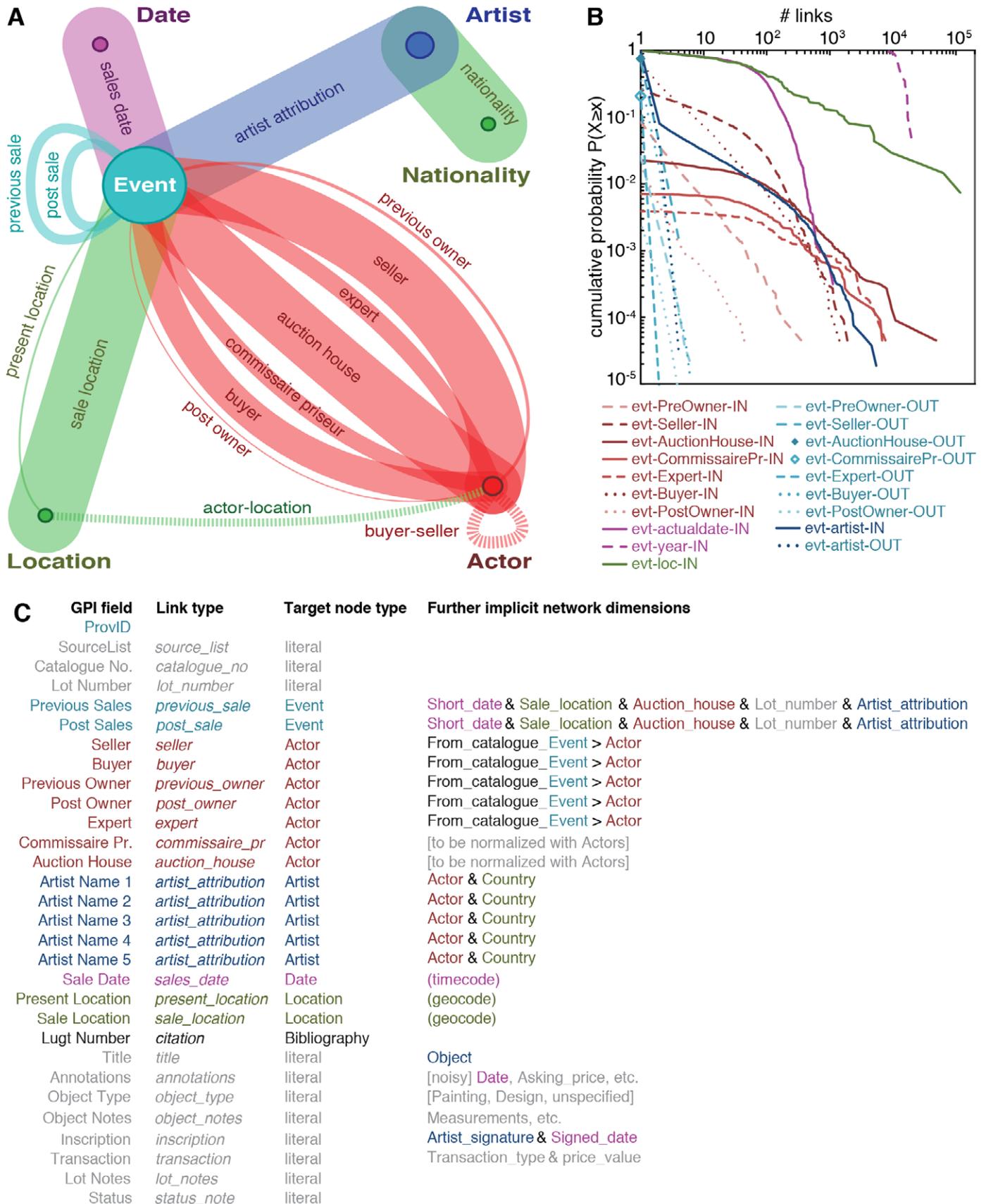

**Fig. 1. The Getty Provenance Index (GPI) is a network of complex networks.** (A) The implicit GPI network data model, equivalent to an entity-relationship-diagram with node size and line width weighted by node and link frequency; (B) The cumulative frequency $P(X \geq x)$ for node IN and OUT degrees, colored by node-type, which establishes the GPI as a subject of network science; (C) The original flat GPI data fields (column 1) with explicated link and target node types (columns 2 & 3), and further implicit network dimensions, which could be explicated in further analysis (column 4).

Figure 1B establishes the GPI as a network of complex networks, i.e. a subject of network science, by summarizing the basic frequency statistics of in- and out-degrees for all link types between sales events connected to dates, locations, artists, and actors. Each data series indicates the cumulative probability $P(X \geq x)$ that nodes of a particular type have at least x connections.[18] Even without deeper quantitative analysis, such as fitting mathematical functions including exponentials or power laws[19], the figure allows for a number of interesting observations. First, all link-types in the plot appear to be asymmetric "quasi-1:n", with a very low, but existing probability for sales events to be connected to multiple actors of the same type, or multiple artist attributions. As a consequence, more sophisticated follow-up analysis would likely make use of network projections[20] or shortcuts between node types, as in case of the buyer-seller network depicted further below. Indeed, Fig. 1B provides interesting hints why such a combination may be of interest. The respective frequency distributions for buyers (evt-Buyer-IN) and sellers (evt-Seller-IN), for example, follow rather different trajectories, with an almost straight power-law-ish distribution for buyers, and a more curved distribution for sellers, both in line with the observation further below that the market seems to broaden, with objects flowing from fewer larger collections to more numerous smaller collections. Finally, Fig. 1B also reveals more curious patterns of art market data, such as the ultra-broad distribution of sales by location (evt-loc-IN), which points to either heavy market concentration to central locations, and/or exceptional under-documentation of the rural art market, when compared to the expected size distribution of cities and towns.[21]

Figure 1C summarizes our "explication" of networks from the original flat GPI data in more detail. With explication we mean the identification of nodes and links from semistructured or unstructured text in the original data files. In principle, the process of explication combines a simple form of entity recognition[22] with graph database normalization[23], where all nodes are assigned with a unique ID and a node type, connected to the original sales record via specific link types. Node and link types are split into axiomatic "dimensions", such as social actors, locations, periods, concepts, etc. The first column in Fig. 1C lists the original GPI data fields, which we have amended with a unique identifier (ProvID) for each sales event. The second and third columns provide the extracted link types and target node types, corresponding to the nodes that we have extracted and normalized as far as possible via the creation of further unique IDs linked to the ProvID, after parsing the content of each original data field. The last column in Fig. 1C, finally, indicates additional network dimensions that could be extracted from the data with further data preparation and processing. Being a proof of concept, our network explication is obviously far from complete, further to include additional dimensions from unstructured text, such as prices or depicted subjects, further to be cleaned, with normalization checked by human editors, and further to be extended to all GPI data, spanning several centuries and many more countries.

### Social dimensions of the GPI

Mentioning the concept of networks in a conversation about provenance or the art market in particular, the first question put on the table is usually the desire to map the network of buyers and sellers. Accordingly, Fig. 2A maps the direct links between the buyers and sellers in the French, British, Belgian, and Dutch art markets from 1801 to 1820 (cf. Fig. S1 for direct actor-location links).[24] The figure was created using minimal amounts of manual optimization after applying a standard spring embedded layout[25], mostly governed by connected actors attracting, and disconnected actors repelling each other. Even while using such a simple network layout, a clear and meaningful structure emerges. In the upper left corner we see the British cluster with an Irish satellite, while mid-center we find the French cluster, both connected to the British and to the more multi-centric clusters of Belgian and Dutch sales. Probably curious for art historians, almost all sales, except for a few in the lower right corner, are indirectly connected to all other sales, within the so-called Giant Connected Component.[26] As such, the European auction market can be legitimately seen as a single integrated system, whose parts cannot be fully understood by studying a single region. While it is true that most actors stay within one region, and stick to either exclusively buying (red) or selling (blue), the figure clearly also shows links between clusters, as well as brokers between communities that both buy and sell (grey). Finally, the node size distribution provides further indication that the market is broadening with many larger, i.e. more frequent sellers (in blue) spreading their artworks to a larger number of much smaller buyers (in red).

A closer look at the individuals connecting multiple marketplaces shows some prominent names such as Jean-Baptiste-Pierre Lebrun or Alexandre-Joseph Paillet. The relative position of an agent in the field is especially telling. While Lebrun overall seems to buy in Paris and sell to London (but also to Rotterdam), Paillet, a much bigger broker, appears more drawn to Belgium. In an interactive representation we could immediately pull out individual agents—some much less known than those just mentioned—and study their behavior over time. Broken down year by year, the network, shows little persistence, as many agents come and go, vanish like stars on the sky (cf. supplementary Fig. S2). However, we detected one dealer who very visibly endured in

---

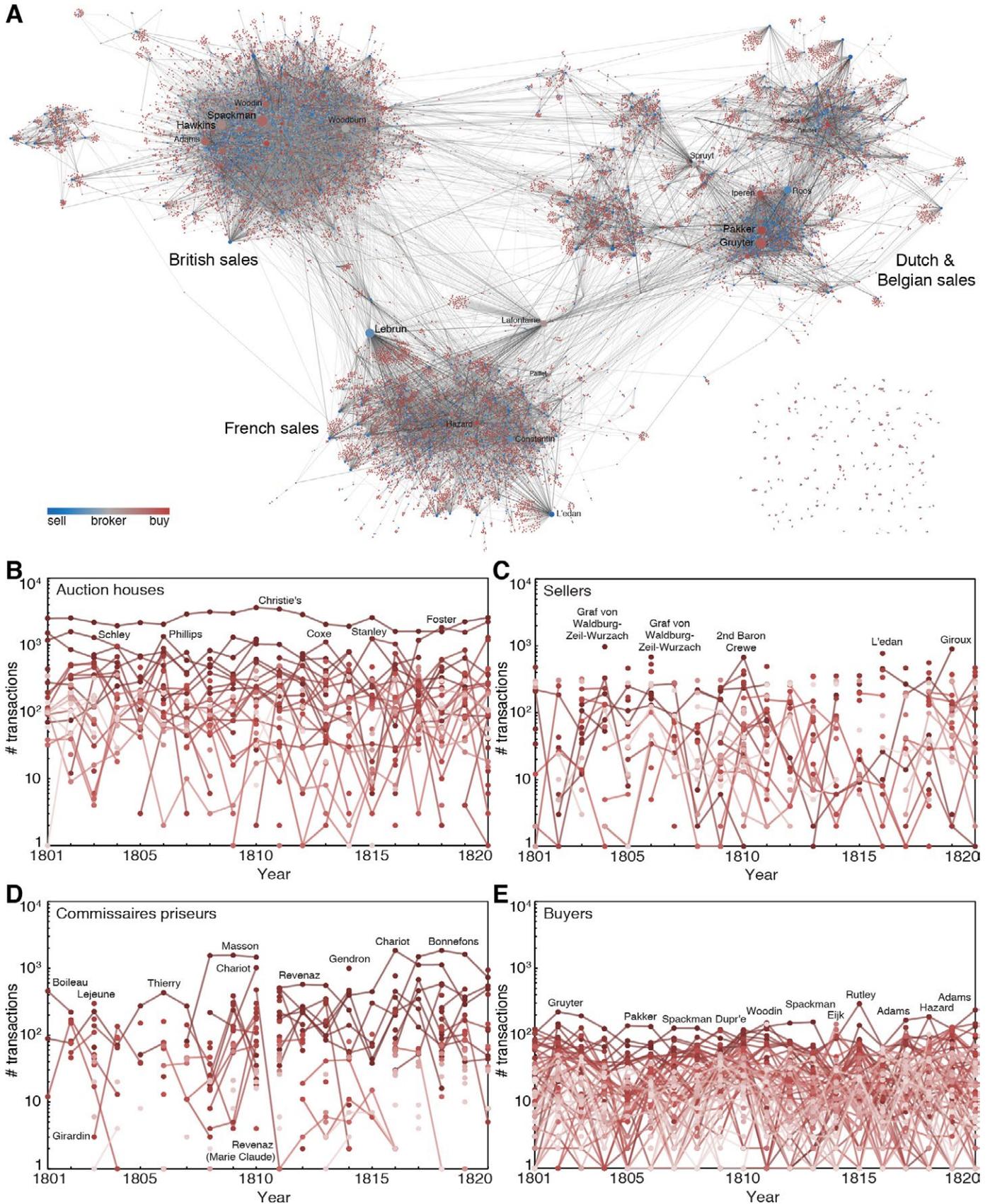

**Fig. 2. GPI social dynamics differ across auction houses, sellers, commissaires priseurs, and buyers.** (A) The GPI social network of predominant sellers (blue), buyers (red), and brokers that buy and sell (grey); node size indicates numbers of sales 1801 to 1820; Clusters correspond to the British (upper left), French (bottom center), and Dutch/Belgian markets (upper right); the yet-to-normalize Belgian/Dutch nodes labeled "Pakker" and "Gruyter"/"Gruiter" indicate that the diagram shows a lower bound of European art market integration; (B) Top auction houses are relatively persistent in the market; (C) Top sellers tend to sell once or twice; (D) Top commissaires priseurs and (E) top buyers typically tend to stay in the market for a number of years. The gap in commissaires priseurs after 1810 is caused by two yet-to-normalize nodes labeled "Revenaz".

the same position over 20 years, connecting all four national art markets. Pierre-Joseph Lafontaine appears overall as a broker (gray), but in the video he changes color over time, being sometimes red, sometimes blue, which means he buys and sells in cycles (cf. supplementary Fig. S2A). Lafontaine, obviously a quite important dealer, has been largely neglected by art history, there is no monographic study so far (a first article on the dealer is in press at Getty publications). When it comes to markets, art history tends to favor a micro-historical approach, devoting monographs to a few international dealers and extrapolating from their behavior.[27] Yet, if we go back to the larger picture, we notice that a majority of the agents in the network operate only locally.

Figures 2B/C/D/E provide a deeper insight into the dynamics of individual social roles from 1801 to 1820. Each plot indicates the number of sales events over time for top individual actors on a logarithmic scale. The plots indicate individual differences in the typical "duration of survival" within the art market system. Auction houses (in Fig. 2B) tend to be active over longer time frames, with Christies being the most "outstanding", while top sellers (in Fig. 2C) tend to appear on the market exactly once, while commissaires priseurs (in Fig. 2D) and top buyers (in Fig. 2E) tend to dominate the market for two to three years. Given these indications, it would certainly be interesting to study the dynamics of ranking within the art market as a whole, similar to the fortunes of cities over centuries or complex systems in general.[28]

**The temporal dimension of the GPI**
At its most granular, the temporal resolution of the GPI is daily, with multi-day sales split into daily lots by Getty editorial convention. As such, an interesting research question pertains to the existence of sense-making temporal patterns and scales, similar to the weekly or bi-weekly pattern of buying groceries. The question is interesting, as the identification of a "natural" timescale in the art market would provide us a basis for further analysis, when looking into other aspects or dimensions over time. In our exploration of the GPI data we have played with temporal resolution, looking at daily, weekly, monthly, and annual resolution, finding an annual pattern that connects our four European markets in a meaningful way. Figure 3A depicts four matrices for the British, French, Dutch and Belgian art markets respectively. Cells in the lines of each matrix indicate the number of sales per month (from low in yellow to high in red) from January to December horizontally, from 1801 to 1820 vertically. Sales per year are summarized on the right border of each matrix, indicating general sales transaction frequency over time. The total number of sales for each month over twenty years is given at the bottom of each matrix. It is interesting to note that the general sales transactions over twenty years are more irregular and specific to each market, even though we find traces of synchronization, such as peaks in sales around 1810 when the volume of annual sales almost doubles in all national market places. The total across months, at least upon first glance on the other hand, is characterized by the same consistent pattern in all four markets, by what appears to be a sine-like oscillation that is shifted in time. Indeed, Fig. 3B superimposes these oscillations in a single plot, and summarizes them further to what looks like a rather smooth oscillation (given in black dots) that peaks in late spring and finds its minimum in the dark months of the year. The most surprising observation is that Dutch and Belgian sales seem to accommodate between the broad peaks of French sales in autumn and British sales in spring. This is interesting as the annual oscillatory behavior is clearly visible in the French and British matrices above, while the oscillation drowns in noise within the Dutch and Belgian matrices, due to the lower number of sales transactions in any particular year. So literally "in sum" the European auction market, even in times of political upheaval between 1801 to 1820, seems to be set up for international integration, allowing for agents to travel between multiple sales locations. The question, if this setup is the result of a memory effect from previous times, or an implicit anticipation, has to be left to future research.[29] More evidence for such integration will emerge from looking at artist attributions and nationality further below. The GPI data for the French, British, Dutch, and Belgian markets from 1801 to 1820 surprisingly contains only 137 sales locations, with the top two locations, London and Paris, concentrating 181,217 of 267,661, i.e. 67.7% of all sales transactions. As mentioned above, this may have two reasons: actual market concentration, or "rural" under-documentation. However, while one may still suspect a major dataset bias in these results, looking closer at the data, we do find great coherence with related phenomena. Figure 3C-F indicate the number of sales transactions per location in the British, French, Dutch, and Belgian markets over time. London (in Fig. 3C) clearly dominates the British market with several thousand sales transactions per year, followed by some significantly smaller locations that are still persistent over years, and many locations that have less than a couple of hundred sales, but none in consecutive years. Paris (in Fig. 3D), even stronger than London, is basically equivalent with the French market, accompanied with only a few side locations, that never see auctions in consecutive years. The Dutch market (in Fig. 3E) is characterized by Amsterdam dominating the market much more weakly, challenged by a number of cities that fluctuate in and out of the market with highly active phases of two to three years. The Belgian market (in Fig. 3F), finally, does not have a center of dominance, but is subject to obvious multicentric competition. All in all, this picture very well reflects the known ranking dynamics of cultural centers in Europe[30], where Paris is subject to a winner-takes-all regime in France, and the Netherlands and Belgium, much like Germany, are subject to a fit-gets-richer regime of multicentric competition. Britain, as expected, is halfway in between. Like with the social role dynamics above, there can be no doubt that a deeper investigation of the ranking dynamics in question would be a very valuable enterprise when applied to longer stretches of art market data. Very recent work provides further evidence towards this point, covering the time-frame between 1985 to 2015, yet to be extended to more historical timescales.[31]

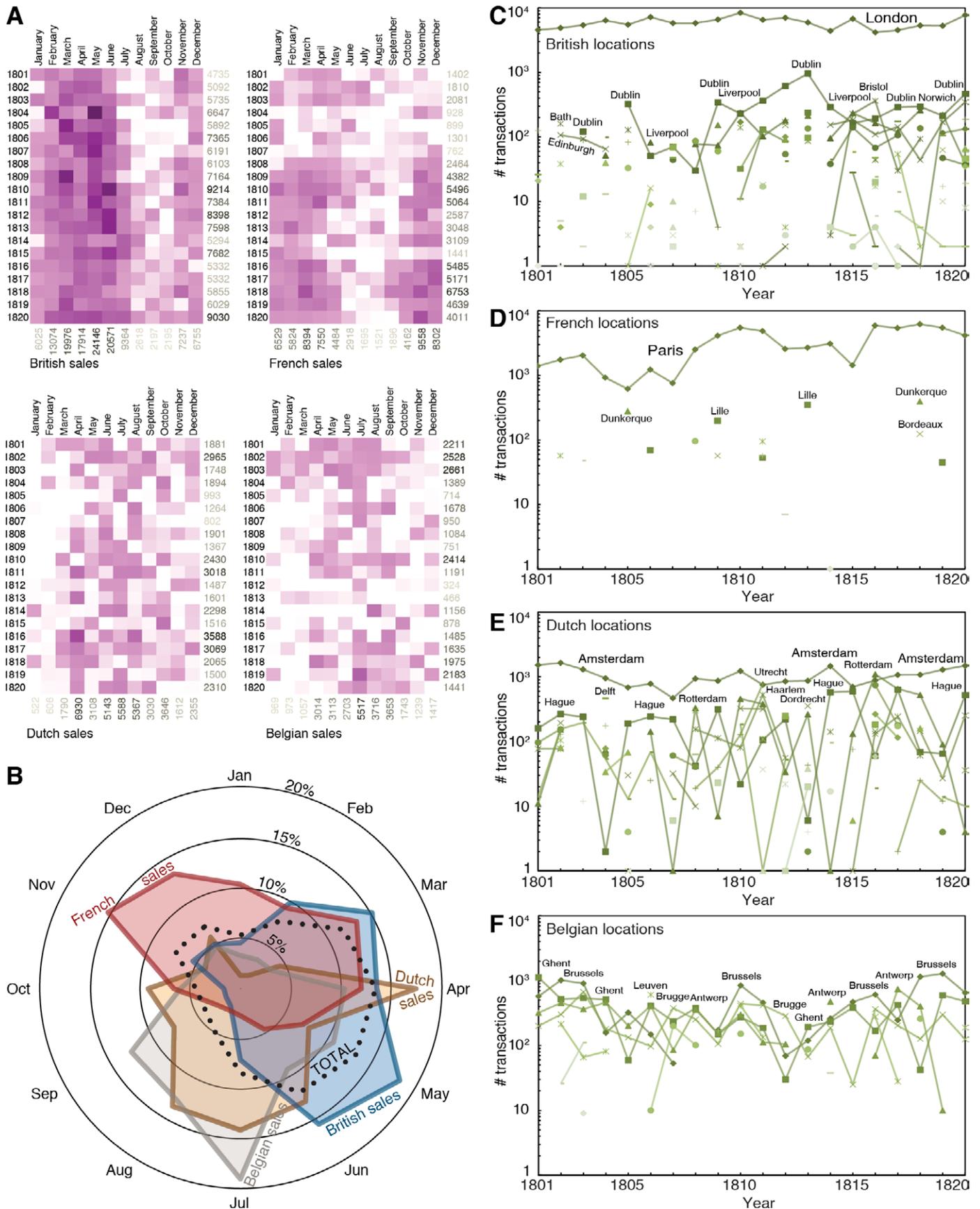

**Fig. 3. GPI spatio-temporal dynamics indicate shifted market cycles and differences in centralization.** (A) Sales frequency over months (x-axis) per year (y-axis) for British, French, Dutch, and Belgian sales summarizes to annual regularity; (B) Market cycles, based on total sales per month for each market, are shifted out of phase, with British sales peaking in May, French in November, and Dutch and Belgian sales accommodating in between; (C-F) Sales frequency distributions for locations in each market over time indicate different regimes of centralization; dots are connected if a location has sales in consecutive years.

**Conceptual dimensions of the GPI**

One of the core missions of provenance research is to trace the trajectory of a given artwork from its creation and its intermediaries to the present owner. Feeding into this mission, the ideal provenance record would allow for unambiguous identification of the artwork in question. Preceding photography, such identification, in particular for common subjects and prolific artists is unfortunately beyond the realm of what can be achieved for the majority of sales transactions. Despite its name, the GPI does not offer long chains of ownership, or, for the same reasons, capture broad documentation of repeated sales of the same object. Based on our analysis of the GPI, we set out to offer an alternative interpretation, that makes meaningful use of the GPI data, even in the majority of cases, when the object of sales cannot be identified with precision. The key premise is that artist attributions need to be understood not so much as precise identifications, but as more or less precise characterizations that function like product categories in a supermarket. Much like "long pasta" is not necessarily "Spaghetti No. 12", the artist attribution "Rubens" does not necessary identify an original, but may indeed just look like Rubens in the eyes of those involved in the sales transaction.

Figure 4A indicates the nationality of the artist attribution associated with sales transaction records in the Dutch, Belgian, French, and British market. While local artworks appear to be unambiguously popular in their respective home markets, Dutch art in particular emerges as the most frequently sold or at least offered in all four markets, easily eclipsing even Italian, German, and Spanish art, which, without any knowledge about art history, one would certainly expect to be more numerous, simply by extrapolating from the population size. Figure 4B, indeed provides a curious explanation that feeds into such an expectation, which from the point of view of an expert art historian may sound naive, as the typical production output of Dutch artists was indeed much higher than that of an Italian master. Fig. 4B plots the percentage of derivative artist attributions in all four markets, with a derivative meaning "school of…", "workshop of…", or "copy after Rembrandt" as opposed to "Rembrandt". One can clearly see that the amount of derivative attributions in the British, French, and Belgian markets from 1801 to 1820 fluctuate more or less bounded between 5% and 15%, while the Dutch derivatives find a low at 21% in 1804, rising to an overwhelming 70% at the end of our two decades.

Curious about this phenomenon of Dutch look-alikes, we decided to probe deeper into artist attributions, i.e. the artist names themselves. As a first approximation, we listed the total frequency of art sales transactions for each artist, to create a top ten list, not based on prices, but purely on frequency of attribution, derivative or not. We assumed the most frequent artists, even if prolific, would not be prolific enough, but instead their names would function as "product categories", i.e. as simple verbal denominators to identify the complicated polymorphic phenomenon that characterizes the sum of works in the style of an artist. Of course, one may note, that all actors involved may still be convinced that the attributions are correct. Figure 4C lists the ten top-ranked, not necessarily authentic images, found via a simple reverse Google Image search by submitting the top-ten most frequent artist attributions in our GPI data.[32] Even without consulting the respective catalog raisonnée, it becomes visibly clear that the top artist attributions in the GPI data work indeed very well as "product categories", representing otherwise complicated-to-describe visual configurations that are easy to disambiguate from another with the bare eye.

Conceptually up to something, while still on thin ice, we decided to test the idea further by employing a standard method used to make sense of modern supermarket and online retail data, a method called market basket analysis, also known as k-itemset analysis.[33] If the artist attributions in the 19th century art market would indeed function like product categories in a supermarket, sure enough, they should exhibit the same statistics as a modern supermarket, given we could identify a sense-making time-frame that amounts to the equivalent of filling the fridge every weekend. Defining the "shopping basket" alternatively as single sales transactions, all transactions of a buyer in a week, in a month, in a year, or in total, we got sense-making results in all cases. Here we report the result for an annual shopping cycle, which is both a bit less arbitrary, based on the annual oscillation of the European auction market identified above in Fig. 3B, and in fact provides the best result in comparison. Figure 4D indicates the cumulative probability $P(X \geq x)$ for the "relative support" or frequency of artist attributions $k=1$, pairs of attributions $k=2$, triples of attributions $k=3$, etc. to appear in a particular "shopping basket". Taking into account the difference in dataset size, the GPI plot in Fig. 4D is very well in line with modern supermarket or online retail data. In sum, even though this may sound provocative or even controversial, we can therefore indeed make the claim that the statistics of the European auction market, or a large portion of it between 1801 and 1820, reveal a system that functions like a super-slow supermarket on an annual grocery-cycle. This result is exciting, as the art market is relatively well documented, and with further digitization, may provide us with a chance to analyze market evolution over very long time-frames, beyond standard methods, such as time-series-analysis of prices, or beyond more anecdotal approaches dealing with particular auction houses, collectors, artists, or artworks.

**Conclusion**

Although, in recent years, reception studies have shifted attention from the heroic artistic producer to the sometimes even anonymous artistic consumer, the grand majority of art historical studies nevertheless focuses on exceptional events and masterpieces. Much has been written, for example, about the move of the famous Orléans collection from Paris to London, where it was dispersed through various private treaty sales and a series of auctions between 1789 and 1802. These prestige sales are often taken as a pars pro toto for the internationalization of the art market in the wake of the French revolution and the emergence of London as its uncontested center. If we "shift the gaze from the extraordinary to the everyday," as Krzysztof Pomian suggested at the peak of the Annales[34], we may well arrive at a different picture, representing a regionally fragmented, but essentially intermingled European art world. After all, the early years of the nineteenth century are characterized by wars and trade restrictions. The ability to redirect attention and pose new questions is the strong suit of data-driven and computational methods. Within the realm of digital art history and cultural analytics, large data sets in conjunction with new methods of visualization allow for a 'distant reading' of the (trans)national circulation of art works, social networks of agents, and cultural consumption more generally. In addition, our results indicate that large art historical data sets, such as the GPI, are valuable in our quest to quantify fundamental patterns and laws, feeding into a systematic science of art and culture. Either complementing or contradicting the predominant case-study approach, shifting our gaze from exceptional events to a large mass of facts will raise a number of new questions we were previously unable to ask.

---

[32] The Google Image Search was performed on June 5, 2014.
[33] Cf. Berry, Michael J., and Gordon Linoff. *Data mining techniques: for marketing, sales, and customer support*. John Wiley & Sons, Inc., 1997; and Kamber, Micheline, Jiawei Han, and Jian Pei. *Data mining: Concepts and techniques*. Elsevier, 2012.
[34] Krzystof Pomian, "L'histoire des structures." in *La nouvelle histoire*, pp. 115-16. Retz, 1978. Quoted in: Moretti, Franco. *Graphs, maps, trees: abstract models for a literary history.* Verso, 2005.

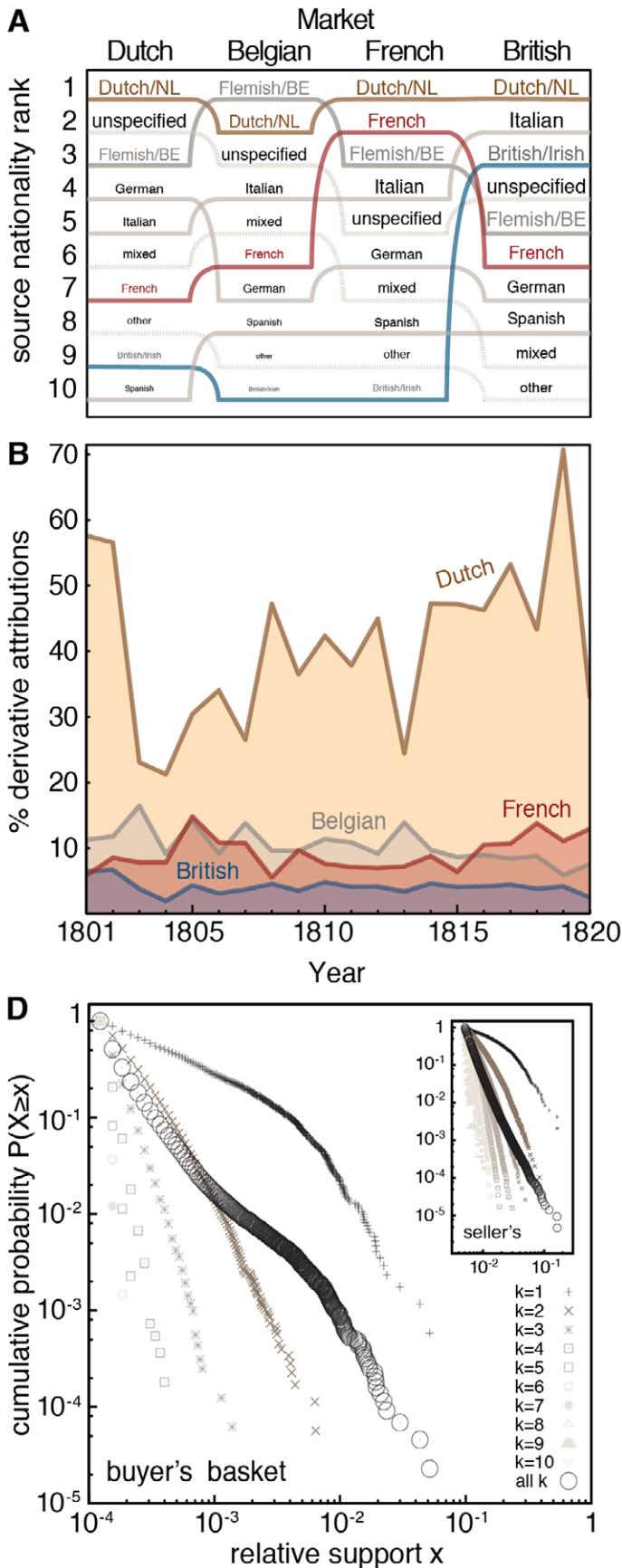
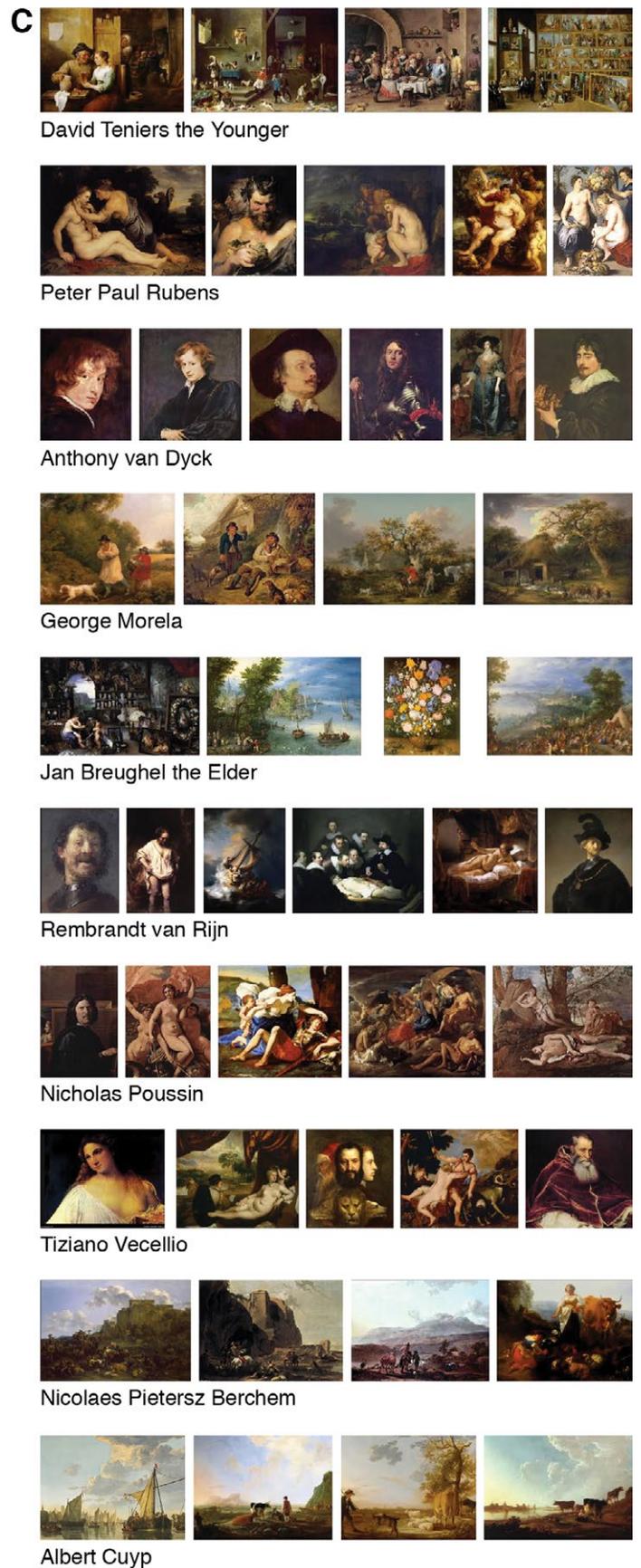

**Fig. 4 GPI artist attributions behave like product categories in a super-slow supermarket.** (A) Top ten source nationalities of artist attributions in each market indicate a general preference for Dutch, Belgian and local art; (B) The fraction of derivative artist attributions over time points to an increasing shortage of Dutch originals; (C) Top Google Image Search results for the top-most-frequent artist attributions are easy to disambiguate by visual family resemblance, obviously with no claim of originality (thumbnails are reproduced here under fair use); (D) The implied product-category-like meaning is confirmed by a frequent-itemset plot, in line with modern market-basket analysis of supermarkets or online-retailers.

**arXiv eprint version; submitted 8 June 2017.**

**Acknowledgements**

Research for this work was performed at BarabásiLab, Northeastern University in 2010, at ETH Zurich in 2012, and the University of Texas at Dallas in 2014 and 2017. The work of MS was partially supported by German Research Foundation (DFG) grant (no. SCHI 1065/2-1), the Chair of Dirk Helbing at ETH Zurich, and the University of Texas at Dallas Arts & Technology (ATEC) Fellowship no. 1. The authors like to thank the Getty Research Institute for providing the data and for initiating and funding the initial meeting.

**Author information**

Maximilian Schich (MS), Christian Huemer (CH), Lev Manovich (LM), Piotr Adamczyk (PA), and Yang-Yu Liu (YL) designed the research; MS prepared data; MS and YL performed research; MS, CH, LM, PA, and YL wrote the manuscript. Correspondance should be addressed to MS, the first author of the paper (maximilian@schich.info).

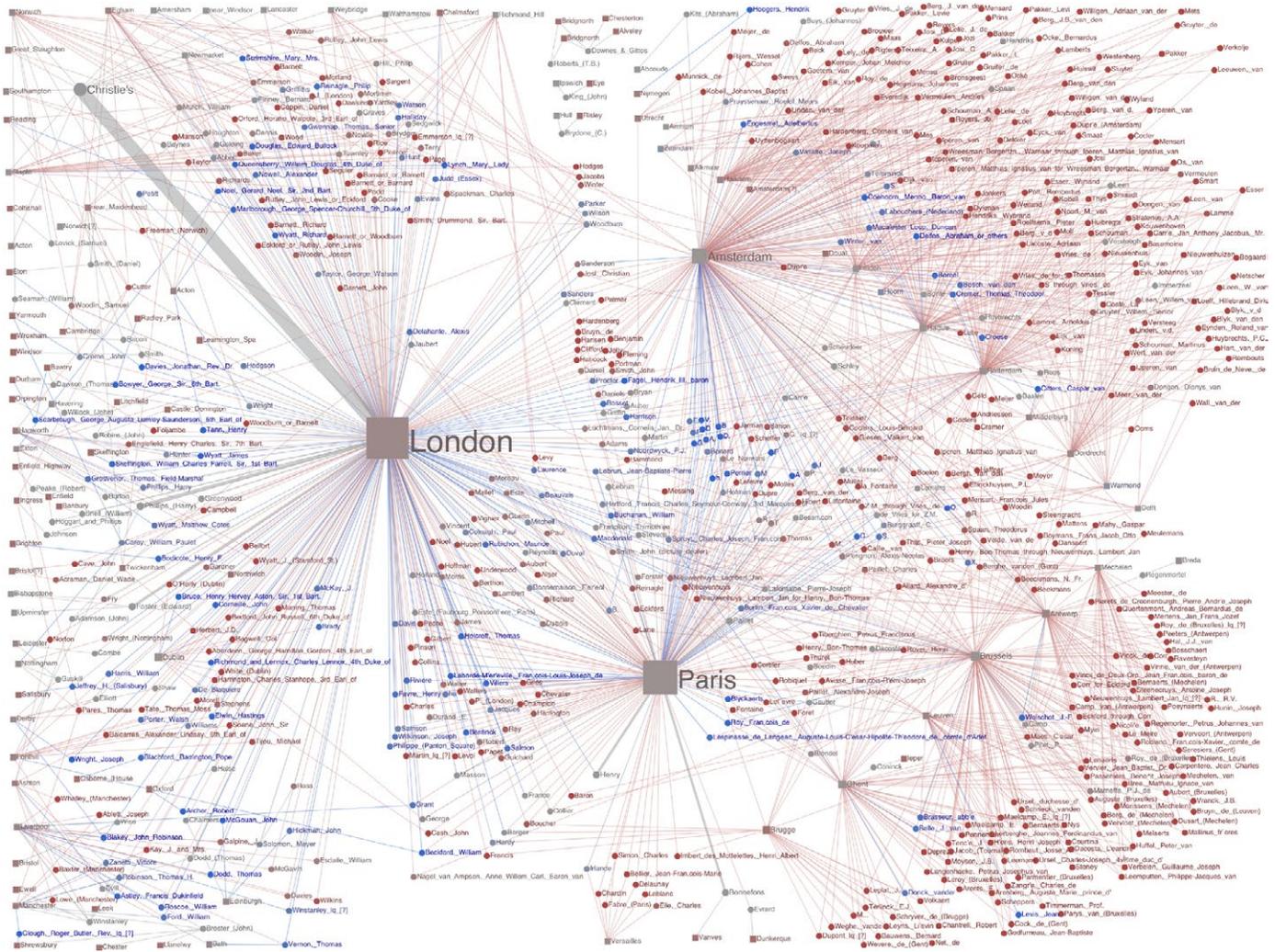

**Fig. S1. The GPI network of social actors connected to at least two sales locations.** The figure indicates the nature of go-between buyers, sellers, autcion houses, commisseures priseurs, and experts. Like in Figure 2A, social actors (circles) and sales locations (squares) are sized by the total number of transactions and colored from predominant selling (blue) to exclusively buying (red), with balanced brokers in between (grey). In the British and French market, go-betweens connecting at least two locations buy, sell, and broker, while the Belgian and Dutch markets are characterized by a majority of predominant buyers tapping into multiple locations. In the center of the picture we find mostly grey brokers that connect all three major theaters (for large zoomable version of this figure, preliminary in terms of color, see http://www.getty.edu/research/tools/provenance/zoomify/index.html).

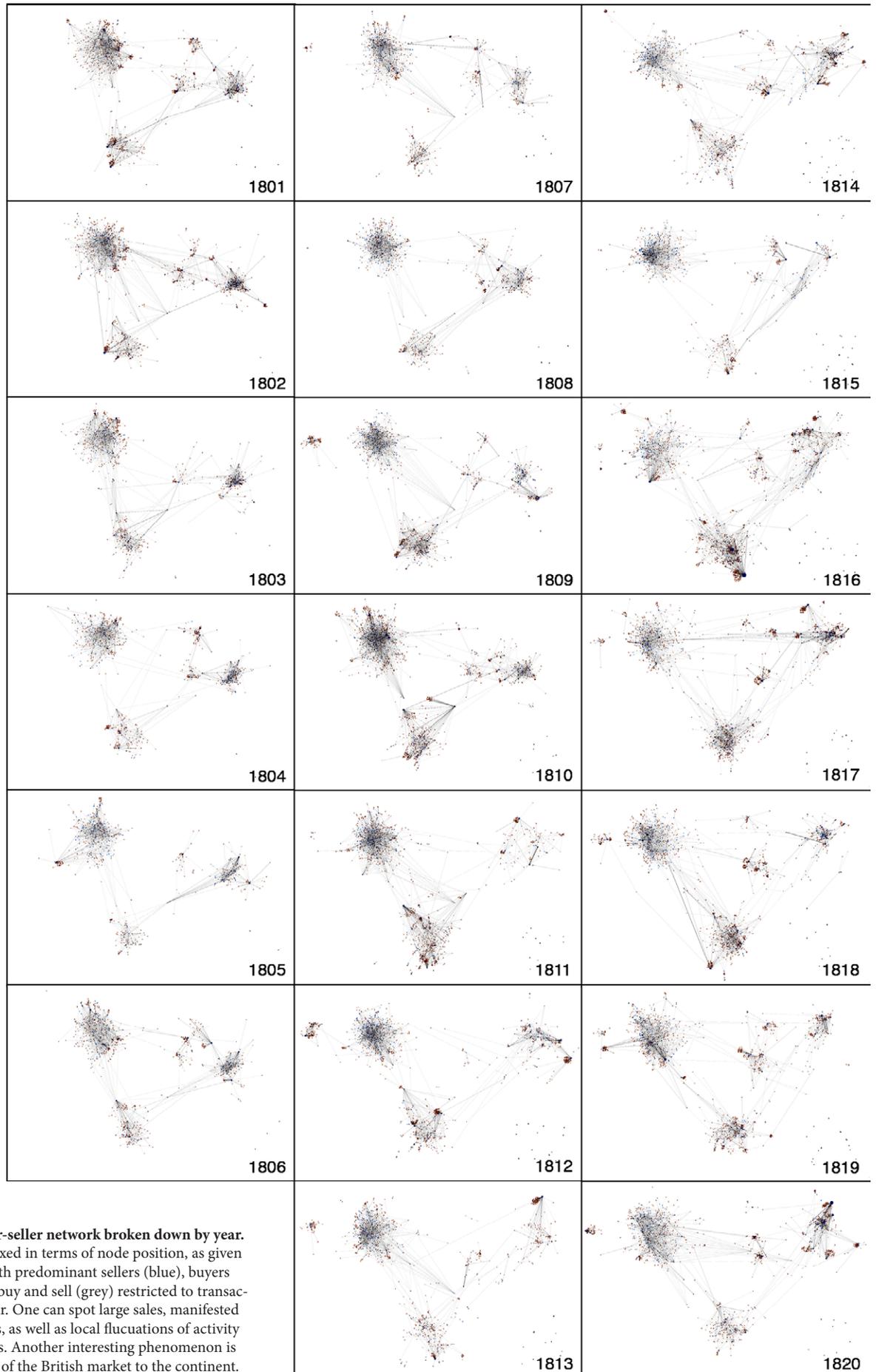

**Fig. S2 . The GPI buyer-seller network broken down by year.** The network layout is fixed in terms of node position, as given in Fig. 2A, only now with predominant sellers (blue), buyers (red), and brokers that buy and sell (grey) restricted to transactions within a given year. One can spot large sales, manifested by brush-like structures, as well as local flucuations of activity in all four market places. Another interesting phenomenon is the varying connection of the British market to the continent.

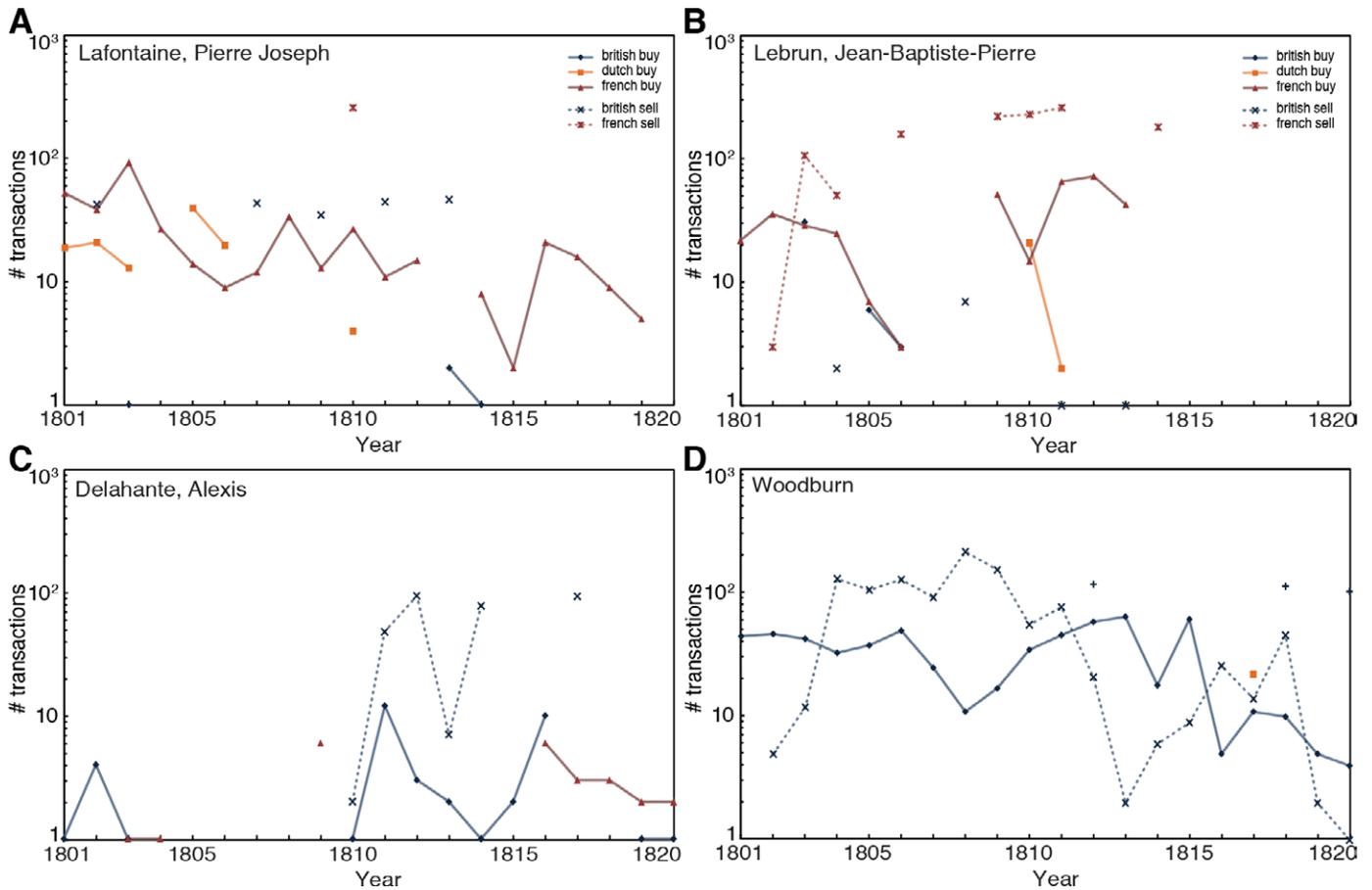

**Fig. S3. The buying and selling activity of selected GPI actors from 1801 to 1820.** (A) Pierre Joseph Lafontaine annually buys in the French and bi-annually sells in the British market over several years; (B) Jean-Baptiste-Pierre Lebrun, (C) Alexis Delahante, and (D) Woodburn exhibit distinctly different trajectories, indicating a rich ecology of actors in the international art market that requires both quantifiaction and qualitative inquiry.